\documentstyle[12pt]{article}
\newlength{\extraspace}
\newlength{\extraspaces}
\newcommand{\be}{\begin{equation}
\addtolength{\abovedisplayskip}{\extraspaces}
\addtolength{\belowdisplayskip}{\extraspaces}
\addtolength{\abovedisplayshortskip}{\extraspace}
\addtolength{\belowdisplayshortskip}{\extraspace}}
\newcommand{\ee}{\end{equation}}

\newcommand{\ba}{\begin{eqnarray}
\addtolength{\abovedisplayskip}{\extraspaces}
\addtolength{\belowdisplayskip}{\extraspaces}
\addtolength{\abovedisplayshortskip}{\extraspace}
\addtolength{\belowdisplayshortskip}{\extraspace}}
\newcommand{\ea}{\end{eqnarray}}

\newcommand{\nonu}{\nonumber \\[.5mm]}
\newcommand{\A}{&\!\!\!}

\begin{document}

\pagenumbering{arabic}

\begin{center}
{\bf Reply  on Brane-World Black Holes and Energy-Momentum Vector}
\end{center}
\centerline{ Gamal G.L. Nashed}

\bigskip

\centerline{{\it Mathematics Department, Faculty of Science, Ain
Shams University, Cairo, Egypt }}

\bigskip
 \centerline{ e-mail:nasshed@asunet.shams.edu.eg}

\hspace{2cm}
\\
\\
\\
\\
\\
\\

{\it We show that the energy distribution of the brane-world black
holes given by Salti et al. in the context of teleparallel theory
is not right. We give the correct formula of energy of those black
holes.}

Salti et  al. \cite{SA} have derived  brane-world  black holes in
the context of tetrad theory of gravitation \cite{SA}. Then,
 calculations of energy in the context of teleparallel geometry
 of those black holes in the {\it spherical polar
coordinate} have been given. The result of these calculations
gives the energy to have the form\footnote{We will use the same
notations given in Ref. \cite{SA}} \be E= \displaystyle{r^2 \over
2}\Im'(r) \sqrt{\displaystyle{\Xi(r) \over r \Im(r)}}. \ee They
\cite{SA} discussed Eq. (1) according to the results of energy of
the same black holes given in the context of {\it general
relativity}. They concluded that the energy distribution of the
brane-world black holes {\it within the context of general
relativity using M\o ller's energy-momentum prescription is the
same as the energy distribution  given in the context of tetrad
theory of gravitation using  M\o ller's energy-momentum complex}.
We have a reply to this discussion:

First of all the calculations of energy {\it within the context of
tetrad theory of gravitation} \cite{Mo1} have been done
in spherical polar coordinate which is not right due to the following reasons:\vspace{0.2cm}\\
1) The energy-momentum vector $P^\mu$ used in this calculation is
not transform as a 4-vector
under linear coordinate transformation \cite{Mo66}.\vspace{0.2cm}\\
2) The superpotential is not invariant under local Lorentz
transformation \cite{AGP}.  So calculations of energy in this
coordinate will not be accurate. The calculations will be more
accurate in the Cartesian coordinate and we have done such
calculations and obtained the necessary components of the
superpotential of Eq. (31) given in Ref.\cite{SA}   \be {{\cal
U}_0}^{0 \alpha}=\displaystyle{ 2 \sqrt{ r \Im(r)} \  n^\alpha
\over \kappa r^2 } \left[\sqrt{\Xi}-\sqrt{r}  \right], \qquad
\alpha=1,2,3,  \qquad n^\alpha=\left(\displaystyle{x \over
r},\displaystyle{y \over r},\displaystyle{z \over r}\right),\ee
using Eq. (39) in Ref. \cite{SA}, the energy distribution of the
brane-world black hole \cite{SA}  is given by \be E(r)= \sqrt{ r
\Im(r)} \left[\sqrt{\Xi}-\sqrt{r} \right], \ee which is different
from Eq. (1). Therefore,  Eqs. (53), (56), (61), (64) and (69) in
Ref. \cite{SA} using Eq. (3) are given by \ba E \A=\A
r\sqrt{{1-\displaystyle{2m \over r} \over 1-\displaystyle{3m \over
2r}}} \left[\sqrt{\left(1-{2m \over r}\right) \left(1-{\lambda_0
\over r}\right)}-\sqrt{1-{3m \over 2r}} \right],  \ for \
\lambda_0=3m/2 \qquad E \cong m, \nonu
E \A=\A \sqrt{r^2-h^2}\left[\sqrt{1-\displaystyle{h^2 \over
r^2}}\left(1+\displaystyle{\chi-h \over \sqrt{2r^2-h^2}}
\right)^{1/2} -1 \right], \nonu
E \A=\A r\left(1-\displaystyle{2m \over r} \right) \left[\sqrt{
\displaystyle{r\left(1-{\lambda_1 \over
r}\right)\left(1-{\lambda_2 \over r}\right)} \over 2}-1\right],
\nonu
E \A=\A \displaystyle{\sqrt{s^2-r^2} \over s^2} \left[
\sqrt{s^2-r^2}\sqrt{r+\displaystyle{Z \over (2s^2-3r^2 )^{3/2}
}}\sqrt{r}-s r \right], \nonu
E \A=\A r\left(1-\displaystyle{2m \over r}\right)^{1/f}
\left[\left(1-\displaystyle{2m \over
r}\right)^{-1}-\left(1-\displaystyle{2m \over
r}\right)^{1/f}\right], \ea respectively. Now we came to the
conclusion that the energy distribution calculated in the context
of tetrad theory of gravitation \cite{Mo1} is different from the
energy distribution  of M\o ller energy-momentum complex
calculated in the context of general relativity theory \cite{LL}.


\begin{thebibliography}{99}

\bibitem{SA} M. Salti and O. Aydogdu, {\it JHEP} 0612 (2006), 078.

\bibitem{Mo1} C. M\o ller, {\it Mat.\ Fys.\ Medd.\ Dan.\ Vid.\ Selsk.\ } {\bf 39} (1978), 13.

\bibitem{Mo66} C.  M\o ller,
 {\it Mat.\ Fys.\ Medd.\ Dan.\ Vid.\ Selsk.\ }{\bf 35} (1966), no.3.

\bibitem{AGP} V.C. de Andrade, L.C.T Guillen  and J.G. Pereira, {\it Phys.\ Rev.\ Lett.\ }
{\bf 84} (2000), 4533; {\it Phys.\ Rev.\ } {\bf D64} (2001),
027502.

\bibitem{LL} L.D. Landau and E.M. Lifshitz, {\rm The Classical Theory of
Fields (Pergamon Press, Oxford, 1980)}.
\end{thebibliography}
\end{document}